\documentclass[11pt]{article}
\usepackage{ejorsj-s2}
\usepackage{amsmath,amssymb,amsfonts,amsthm}
\usepackage{graphicx}
\usepackage{color}
\usepackage{bm,booktabs,url} 
\eqswitchon

\title{PIECEWISE-LINEAR APPROXIMATION FOR \\ FEATURE SUBSET SELECTION IN A SEQUENTIAL LOGIT MODEL}
\author{
\begin{tabular}[h]{ccc}  
Toshiki Sato & Yuichi Takano & Ryuhei Miyashiro \\ 
\textit{University of Tsukuba} & \textit{Senshu University} & \textit{Tokyo University of Agriculture and Technology} \\ 
\end{tabular}
}
\date{(October 7, 2015)}
\begin{document}
\maketitle
\begin{abstract}
This paper concerns a method of selecting a subset of features for a sequential logit model. 
Tanaka and Nakagawa~(2014) proposed a mixed integer quadratic optimization formulation for solving the problem based on a quadratic approximation of the logistic loss function. 
However, since there is a significant gap between the logistic loss function and its quadratic approximation, their formulation may fail to find a good subset of features. 
To overcome this drawback, we apply a piecewise-linear approximation to the logistic loss function. 
Accordingly, we frame the feature subset selection problem of minimizing an information criterion as a mixed integer linear optimization problem. 
The computational results demonstrate that our piecewise-linear approximation approach found a better subset of features than the quadratic approximation approach. 
\end{abstract}
\section{Introduction}

The analysis of ordinal categorical data~\cite{Ag10,LiAg05} is required in various areas including finance, econometrics, and bioinformatics. 
For instance, credit ratings of financial instruments are typical ordinal categorical data, and thus, many previous studies have analyzed such data by means of ordinal classification models (see, e.g., \cite{AfGo09,Ed10,PoFi99}). 
A sequential logit model~\cite{Am81,KaMo79,Tu91}, also known as the continuation ratio model~\cite{Ag90,Fi80}, is a commonly used ordinal classification model. 
It predicts an ordinal class label for each sample by successively applying separate logistic regression models. 
One can find various applications of sequential logit models: Kahn and Morimune~\cite{KaMo79} used this model to explain the duration of unemployment of workers; Weiler~\cite{We86} investigated the choice behavior of potential attendees in higher education institutions; Fu and Wilmot~\cite{FuWi04} estimated dynamic travel demand caused by hurricane evacuation. 

In order to enhance the reliability of these data analyses, it is critical to carefully choose a set of relevant features for model construction. 
Such a feature subset selection problem is of essential importance in statistics, data mining, artificial intelligence, and machine learning (see, e.g., \cite{BlLa97,GuEl03,KoJo97,Mi02}). 
The mixed integer optimization (MIO) approach to feature subset selection has recently received a lot of attention as a result of algorithmic advances and hardware improvements (see, e.g., \cite{BeKi15,KoTa10,KoYa09,MiTa15a,MiTa15b}). 
In contrast to heuristic algorithms, e.g., stepwise regression~\cite{Ef60}, $L_1$-penalized regression~\cite{Ar12,Ki08}, and metaheuristic strategies~\cite{Yu09}, the MIO approach has the potential of providing an optimality guarantee for the selected set of features under a given goodness-of-fit measure.

Tanaka and Nakagawa~\cite{TaNa14} recently devised an MIO formulation for feature subset selection in a sequential logit model. 
It is hard to exactly solve the feature subset selection problem for a sequential logit model, because its objective contains a nonlinear function called the logistic loss function. 
To resolve this issue, they employed a quadratic approximation of the logistic loss function. 
The resultant mixed integer quadratic optimization (MIQO) problem can be solved with standard mathematical optimization software; however, there is a significant gap between the logistic loss function and its quadratic approximation. 
As a result, the MIQO formulation may fail to find a good-quality solution to the original feature subset selection problem. 

The purpose of this paper is to give a novel MIO formulation for feature subset selection in a sequential logit model. 
Sato et al.~\cite{SaTa15} used a piecewise-linear approximation in the feature subset selection problem for binary classification. 
In line with Sato et al.~\cite{SaTa15}, we shall apply a piecewise-linear approximation to the sequential logit model for ordinal multi-class classification. 
Consequently, the problem is posed as a mixed integer linear optimization (MILO) problem. 
This approach is capable of approximating the logistic loss function more accurately than the quadratic approximation does. 
Moreover, our MILO formulation has the advantage of selecting a set of features with an optimality guarantee on the basis of an information criterion, such as the Akaike information criterion (AIC, \cite{Ak74}) or Bayesian information criterion (BIC, \cite{Sc78}). 

The effectiveness of our MILO formulation is assessed through computational experiments on several datasets from the UCI Machine Learning Repository~\cite{BaLi13}. 
The computational results demonstrate that our piecewise-linear approximation approach found a better subset of features than the quadratic approximation approach did. 

\section{Sequential Logit Model}

Let us suppose that we are given $n$ samples of pairs, $(\bm{x}_i,y_i)$ for $i=1,2,\ldots,n$. 
Here, $\bm{x}_i=(x_{i1},x_{i2},\ldots,x_{ip})^{\top}$ is a $p$-dimensional feature vector, and $y_i \in \{1,2,\ldots,m+1\}$ is a ordinal class label to be predicted for each sample $i=1,2,\ldots,n$. 
In the sequential logit model for ordinal classification, we sequentially apply the following logistic regression models in order to predict a class label of each sample (see, e.g., \cite{Am81,KaMo79,Tu91}), 
\begin{align}
q_k(\bm{x})=\mathrm{Pr}(y=k\mid y\ge k,\bm{x})=\frac{1}{1+\exp(-(\bm{w}_k^\top\bm{x}+b_k))}\quad (k=1,2,\ldots,m),
\label{eq:logreg}
\end{align}
where the intercept, $b_k$, and the $p$-dimensional coefficient vector, $\bm{w}_k=(w_{1k},w_{2k},\ldots,w_{pk})^{\top}$, are parameters to be estimated. 

As shown in Figure~\ref{fig:seqlog}, a feature vector $\bm{x}$ is moved into class 1 with a probability $q_1(\bm{x})$. 
In the next step, it falls into class 2 with a probability $(1 - q_1(\bm{x}))q_2(\bm{x})$. 
In the similar manner, it reaches class $k$ with a probability $(1 - q_1(\bm{x}))(1 - q_2(\bm{x})) \cdots (1 - q_{k-1}(\bm{x})) q_{k}(\bm{x})$. 

\setlength\unitlength{1pt}
\begin{figure}[!t]
\centering
\begin{picture}(400,160)(0,0)
\put(0,0){\line(1,0){400}}
\put(0,160){\line(1,0){400}}

\put(10,120){\vector(1,0){60}}
\put(80,120){\vector(1,0){60}}
\put(150,120){\vector(1,0){60}}
\put(250,120){\vector(1,0){60}}
\put(320,120){\vector(1,0){60}}

\put(10,120){\vector(1,-1){63}}
\put(80,120){\vector(1,-1){63}}
\put(150,120){\vector(1,-1){63}}
\put(250,120){\vector(1,-1){63}}
\put(320,120){\vector(1,-1){63}}

\put(20,130){{\scriptsize$1-q_1(\bm{x})$}}
\put(90,130){{\scriptsize$1-q_2(\bm{x})$}}
\put(160,130){{\scriptsize$1-q_3(\bm{x})$}}
\put(260,130){{\scriptsize$1-q_{m-1}(\bm{x})$}}
\put(340,130){{\scriptsize$1-q_m(\bm{x})$}}

\put(25,70){{\scriptsize$q_1(\bm{x})$}}
\put(95,70){{\scriptsize$q_2(\bm{x})$}}
\put(165,70){{\scriptsize$q_3(\bm{x})$}}
\put(260,70){{\scriptsize$q_{m-1}(\bm{x})$}}
\put(340,70){{\scriptsize$q_m(\bm{x})$}}

\put(10,120){\color{white}{\circle*{15}}}
\put(80,120){\color{white}{\circle*{15}}}
\put(150,120){\color{white}{\circle*{15}}}
\put(250,120){\color{white}{\circle*{15}}}
\put(320,120){\color{white}{\circle*{15}}}
\put(390,120){\color{white}{\circle*{15}}}

\put(80,50){\color{white}{\circle*{15}}}
\put(150,50){\color{white}{\circle*{15}}}
\put(320,50){\color{white}{\circle*{15}}}
\put(390,50){\color{white}{\circle*{15}}}

\put(10,120){\circle{15}}
\put(80,120){\circle{15}}
\put(150,120){\circle{15}}
\put(320,120){\circle{15}}
\put(390,120){\circle{15}}

\put(80,50){\circle{15}}
\put(150,50){\circle{15}}
\put(320,50){\circle{15}}
\put(390,50){\circle{15}}

\put(70,30){{\scriptsize$y=1$}}
\put(140,30){{\scriptsize$y=2$}}
\put(300,30){{\scriptsize$y=m-1$}}
\put(375,30){{\scriptsize$y=m$}}
\put(370,100){{\scriptsize$y=m+1$}}

\put(230,80){$\cdots$}
\end{picture}
\caption{Diagram of sequential logit model}
\label{fig:seqlog}
\end{figure}
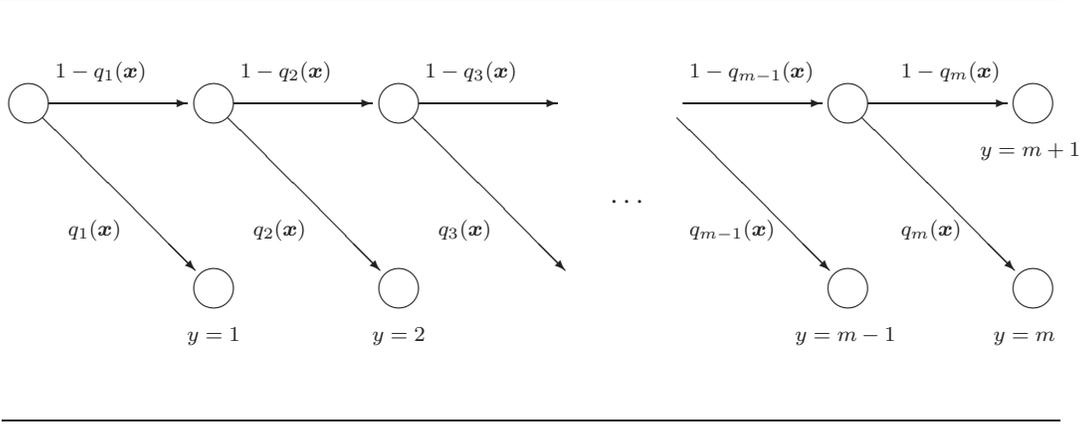

Here, we define
\begin{align}
\delta_{ik}=\left\{\begin{array}{ll}1 & \mathrm{if~}y_i=k, \\ 0 & \mbox{otherwise},\end{array}\right. \qquad (i=1,2,\ldots,n;k=1,2,\ldots,m).
\label{eq:ind1}
\end{align}
It then follows that 
\begin{align}
1-\sum_{j=1}^{k}\delta_{ij}
=\left\{\begin{array}{ll}1 & \mathrm{if~}k<y_i, \\ 0 & \mathrm{otherwise},\end{array}\right.\quad (i=1,2,\ldots,n;k=1,2,\ldots,m).
\label{eq:ind2}
\end{align}
Therefore, the occurrence probability of a sample $(\bm{x}_i,y_i)$ is modeled as follows: 
\begin{align}
\prod_{k=1}^{m}\Big(1-q_k(\bm{x}_i)\Big)^{1-\sum_{j=1}^{k}\delta_{ij}}\Big(q_k(\bm{x}_i)\Big)^{\delta_{ik}} \quad (i=1,2,\ldots,n). 
\label{eq:seqlog}
\end{align}

We refer to model~\eqref{eq:seqlog} as a forward sequential logit model because binary classification models~\eqref{eq:logreg} are applied in order from $k=1$ to $k=m$. 
We can also consider the backward model that makes binary classification in the reverse order from $k=m$ to $k=1$. 
It is known that these two models do not produce the same results~(see \cite{TaNa14}). 

The maximum likelihood estimation method estimates the parameters, $\bm{b}=(b_1,b_2,\ldots,b_m)^{\top}$ and $\bm{W}=(\bm{w}_1,\bm{w}_2,\ldots,\bm{w}_m)$, so that the log likelihood function, $L(\bm{b},\bm{W})$, is maximized:
\begin{align}
 &L(\bm{b},\bm{W})=\log\prod_{i=1}^{n}\prod_{k=1}^{m}\Big(1-q_k(\bm{x}_i)\Big)^{1-\sum_{j=1}^{k}\delta_{ij}}\Big(q_k(\bm{x}_i)\Big)^{\delta_{ik}}  \notag \\
=&\sum_{i=1}^{n}\sum_{k=1}^{m}\left(\left(1-\sum_{j=1}^{k}\delta_{ij}\right)\log(1-q_k(\bm{x}_i))+\delta_{ik}\log(q_k(\bm{x}_i))\right) \notag \\
=&\sum_{i=1}^{n}\sum_{k=1}^{m}\left(\left(1-\sum_{j=1}^{k}\delta_{ij}\right)\log\left(\frac{1}{1+\exp(\bm{w}_k^\top\bm{x}+b_k)}\right)+\delta_{ik}\log\left(\frac{1}{1+\exp(-(\bm{w}_k^\top\bm{x}+b_k))}\right)\right) \notag \\
=&-\sum_{i=1}^{n}\left(\sum_{k=1}^{m}\left(1-\sum_{j=1}^{k}\delta_{ij}\right)f(-(\bm{w}_k^\top\bm{x}_i+b_k))+\sum_{k=1}^{m}\delta_{ik}f(\bm{w}_k^\top\bm{x}_i+b_k)\right), \label{eq:loglike1}
\end{align}
where 
\begin{align}
f(v) = \log(1+\exp(-v)).
\label{eq:logloss}
\end{align}
The function $f(v)$ is called the logistic loss function. 
This function is convex because its second derivative always has a positive value. 
Hence, maximizing the log likelihood function~\eqref{eq:loglike1} is a convex optimization problem. 

From \eqref{eq:ind1}, \eqref{eq:ind2} and \eqref{eq:loglike1}, we obtain a compact formulation of the log likelihood function,
\begin{align}
L(\bm{b},\bm{W})
&=-\sum_{i=1}^{n}\left(\sum_{k=1}^{y_i-1}f(-(\bm{w}_k^\top\bm{x}_i+b_k))+\sum_{k=1}^{m}\delta_{ik}f(\bm{w}_k^\top\bm{x}_i+b_k)\right) \notag \\
&=-\sum_{i=1}^{n}\sum_{k=1}^{m} |\psi_{ik}|f(\psi_{ik}(\bm{w}_k^\top\bm{x}_i+b_k)),
\notag
\end{align}
where
\begin{align}
\psi_{ik}=\left\{
\begin{array}{rl}
-1 & \mbox{if}~k < y_i, \\
1 & \mbox{if}~k = y_i, \\
0  & \mbox{otherwise}, \\
\end{array}
\right.
\quad (i=1,2,\ldots,n;k=1,2,\ldots,m). \notag
\end{align}

\section{Mixed Integer Optimization Formulations for Feature Subset Selection}

This section presents mixed integer optimization (MIO) formulations for feature subset selection in the sequential logit model. 

\subsection{Mixed integer nonlinear optimization formulation}
Similarly to the previous research~\cite{Ar12,Ka09,NaKo09,Tu10}, we shall employ information criteria, e.g., the Akaike information criterion (AIC,~\cite{Ak74}) and Bayesian information criterion (BIC,~\cite{Sc78}), as a goodness-of-fit measure for the sequential logit model. 

Let $S \subseteq \{1,2,\ldots,p\}$ be a set of selected features. 
Accordingly, by setting the coefficients of other candidate features to zero, most information criteria can be expressed as follows:
\begin{align}
-2 \max \{L(\bm{b},\bm{W}) \mid w_{jk} = 0~(j \not\in S;k=1,2,\ldots,m)\} + Fm(|S|+1), 
\label{eq:infcri}
\end{align}
where $F$ is a penalty for the number of selected features. 
For instance, $F = 2$ and $F = \log(n)$ correspond to the AIC and BIC, respectively.

Let $\bm{z}=(z_1,z_2,\ldots,z_p)^{\top}$ be a vector of 0-1 decision variables; $z_j = 1$ if $j \in S$; $z_j = 0$, otherwise. 
The feature subset selection problem for minimizing the information criterion~\eqref{eq:infcri} of the sequential logit model can be formulated as a mixed integer nonlinear optimization (MINLO) problem, 
\begin{align}
\underset{\bm{b},\bm{W},\bm{z}}{\mathrm{minimize}} & \quad 2\sum_{i=1}^{n}\sum_{k=1}^{m}|\psi_{ik}|f(\psi_{ik}(\bm{w}_k^\top\bm{x}_i+b_k))+Fm\left(\sum_{j=1}^{p}z_j +1\right) \label{MINLO:obj} \\
\mathrm{subject~to}
& \quad z_j=0\Rightarrow w_{jk}=0 \quad (j=1,2,\ldots,p;k=1,2,\ldots,m), \label{MINLO:const1} \\
& \quad z_j\in\{0,1\} \quad (j=1,2,\ldots,p). \label{MINLO:const2}
\end{align}

The logical implications~\eqref{MINLO:const1} can be represented by a special ordered set type one (SOS1) constraint~\cite{Be63,BeTo70}. 
This constraint implies that not more than one element in the set can have a nonzero value, and it is supported by standard MIO software. 
Therefore, to incorporate the logical implications~\eqref{MINLO:const1}, it is only necessary to impose the SOS1 constraint on $\{1 - z_j, w_{jk}\}~(j=1,2,\ldots,p;k=1,2,\ldots,m)$. 
Indeed, if $z_j = 0$, then $1 - z_j$ has a nonzero value, and $w_{jk}$ must be zero from the SOS1 constraints. 

\subsection{Quadratic approximation}
The objective function~\eqref{MINLO:obj} to be minimized is a convex but nonlinear function, which may cause numerical instabilities in the computation. 
Moreover, most MIO software cannot handle such a nonlinear objective function. 
In view of these facts, Tanaka and Nakagawa~\cite{TaNa14} used a quadratic approximation of the logistic loss function.

The second-order Maclaurin series of the logistic loss function~\eqref{eq:logloss} is written as follows:
\begin{align}
f(v) \approx \frac{v^2}{8} -\frac{v}{2}+\log 2. 
\label{eq:2ms}
\end{align}
This quadratic approximation of the logistic loss function reduces the MINLO problem~\eqref{MINLO:obj}--\eqref{MINLO:const2} to a mixed integer quadratic optimization (MIQO) problem, 
\begin{align}
\underset{\bm{b},\bm{W},\bm{z}}{\mathrm{minimize}} & \quad 2\sum_{i=1}^{n}\sum_{k=1}^{m}|\psi_{ik}|\left( \frac{\psi_{ik}^2 (\bm{w}_k^\top\bm{x}_i+b_k)^2}{8} -\frac{\psi_{ik}(\bm{w}_k^\top\bm{x}_i+b_k)}{2}+\log 2 \right) \notag \\ 
& \qquad \qquad +Fm\left(\sum_{j=1}^{p}z_j +1\right) \label{MIQO:obj} \\
\mathrm{subject~to}
& \quad z_j=0\Rightarrow w_{jk}=0 \quad (j=1,2,\ldots,p;k=1,2,\ldots,m), \label{MIQO:const1} \\
& \quad z_j\in\{0,1\} \quad (j=1,2,\ldots,p). \label{MIQO:const3}
\end{align}

\begin{figure}[!t]
	\begin{center}
	\includegraphics[scale=0.5]{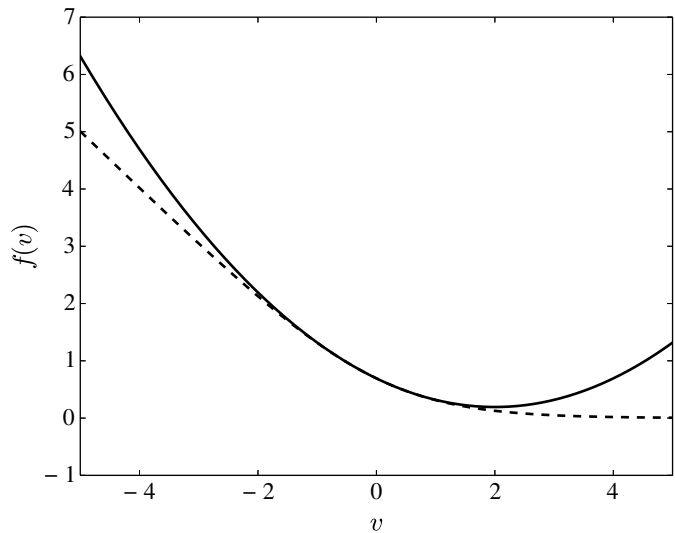}
	\caption{Logistic loss function and its quadratic approximation}
	\label{fig:qa}
	\end{center}
\end{figure}

Figure~\ref{fig:qa} shows the graphs of the logistic loss function~\eqref{eq:logloss} (dashed curve) and its quadratic approximation~\eqref{eq:2ms} (solid curve). 
We can see that the approximation error sharply increases with distance from $v = 0$. 
More importantly, the quadratic approximation function increases on the right side, while the logistic loss function monotonically decreases so that it reduces penalties on correctly classified samples. 
This means that the quadratic approximation imposes large penalties on correctly classified samples. 
Consequently, the MIQO problem~\eqref{MIQO:obj}--\eqref{MIQO:const3} may fail to find a good subset of features. 

\subsection{Piecewise-linear approximation}
In order to approximate the logistic loss function more accurately, we propose the use of a piecewise-linear approximation instead of a quadratic approximation. 

By following Sato et al.~\cite{SaTa15}, we make a piecewise-linear approximation of the logistic loss function. 
Let $V = \{v_1,v_2,\ldots,v_h\}$ be a set of $h$ discrete points. 
Since the graph of a convex function lies above its tangent lines, the logistic loss function~\eqref{eq:logloss} can be approximated by the pointwise maximum of a family of tangent lines, that is, 
\begin{align}
f(v) & \approx \max\{f'(v_{\ell})(v - v_{\ell}) + f(v_{\ell}) \mid \ell = 1,2,\ldots,h \} \notag \\
& = \min\{t \mid t \ge f'(v_{\ell})(v - v_{\ell}) + f(v_{\ell}) \quad (\ell=1,2,\ldots,h) \}. \notag
\end{align}

\begin{figure}[t]
	\begin{center}
	\includegraphics[scale=0.5]{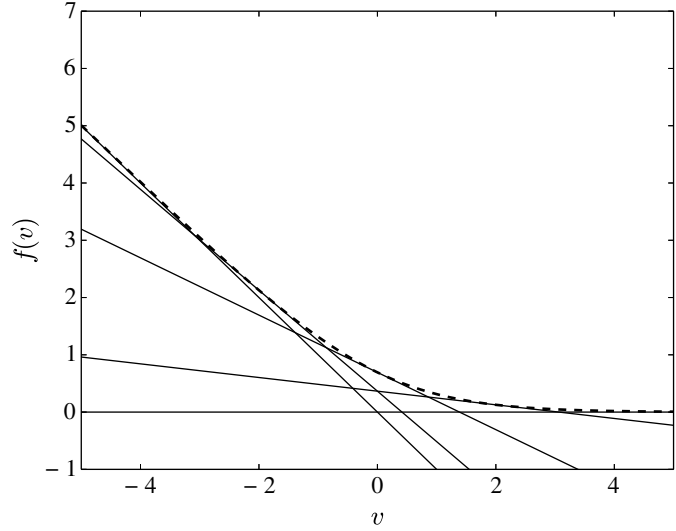}
	\caption{Logistic loss function and its tangent lines}
	\label{fig:pla}
	\end{center}
\end{figure}

Figure~\ref{fig:pla} shows the graph of the logistic loss function~\eqref{eq:logloss} (dashed curve) together with the tangent lines (solid lines) at $v_1 = -\infty, v_2 = -1, v_3 = 1, \mbox{and}~v_4 = \infty$. 
Also note that 
\begin{align}
& f'(v_1)(v - v_1) + f(v_1) = -v, \notag \\
& f'(v_4)(v - v_4) + f(v_4) = 0. \notag 
\end{align}
As shown in Figure~\ref{fig:pla}, the pointwise maximum of the four tangent lines creates a piecewise-linear underestimator of the logistic loss function. 
It is clear that this approach approximates the logistic loss function more accurately than the quadratic approximation approach does. 

By utilizing a piecewise-linear approximation of the logistic loss function, the feature subset selection problem for the sequential logit model can be posed as a mixed integer linear optimization (MILO) problem, 
\begin{align}
\underset{\bm{b},\bm{T},\bm{W},\bm{z}}{\mathrm{minimize}} & \quad 2\sum_{i=1}^{n}\sum_{k=1}^{m}|\psi_{ik}|t_{ik}+Fm\left(\sum_{j=1}^{p}z_j +1\right) \label{MILO:obj} \\
\mathrm{subject~to}
& \quad t_{ik}\ge f'(v_\ell)(\psi_{ik}(\bm{w}_k^\top\bm{x}_i+b_k)-v_\ell)+f(v_\ell) \notag \\
& \qquad \qquad (i=1,2,\ldots,n;k=1,2,\ldots,m;\ell=1,2,\ldots,h), \label{MILO:const1} \\
& \quad z_j=0\Rightarrow w_{jk}=0 \quad (j=1,2,\ldots,p;k=1,2,\ldots,m), \label{MILO:const2} \\
& \quad z_j\in\{0,1\} \quad (j=1,2,\ldots,p), \label{MILO:const3}
\end{align}
where $\bm{T}=(t_{ik};i=1,2,\ldots,n,k=1,2,\ldots,m)$ is a decision variable for calculating the value of piecewise-linear approximation function. 

This MILO problem approaches the original MINLO problem~\eqref{MINLO:obj}--\eqref{MINLO:const2} by increasing the number of tangent lines at appropriate points. 
Moreover, this MILO problem, as well as the MIQO problem~\eqref{MIQO:obj}--\eqref{MIQO:const3}, can be solved with standard mathematical optimization software. 

\section{Computational Results}

This section compares the effectiveness of our piecewise-linear approximation approach with that of the quadratic approximation approach employed by Tanaka and Nakagawa~\cite{TaNa14}. 

We downloaded eight datasets for ordinal classification from the UCI Machine Learning Repository~\cite{BaLi13}. 
Table~\ref{tab:ins} lists these instances, where $n$ and $p$ are the number of samples and number of candidate features, respectively; and \#class is the number of ordinal class labels, i.e., $m+1$. 

\begin{table}[!t]
\centering
\caption{List of instances}
\begin{tabular}{lrrrl}
\toprule
abbreviation & $n$ & $p$ & \#class & original dataset~\cite{BaLi13} \\ \hline
\texttt{Wine-R}& 1599& 11 & 6 & Wine Quality (red wine) \\
\texttt{Wine-W}& 4898& 11 & 7 & Wine Quality (white wine) \\
\texttt{Skill} & 3338& 18 & 7 & SkillCraft1 Master Table Dataset \\
\texttt{Choice} & 1474 & 21 & 3 & Contraceptive Method Choice \\
\texttt{Tnns-W} & 118  & 31 & 7 & Tennis Major Tournament (Wimbledon-women) \\
\texttt{Tnns-M} & 113  & 33 & 7 & Tennis Major Tournament (Wimbledon-men) \\ 
\texttt{Stdnt-M} & 395  & 40 & 18 & Student Performance (mathematics) \\ 
\texttt{Stdnt-P} & 649  & 40 & 17 & Student Performance (Portuguese language) \\
\bottomrule
\end{tabular}
\label{tab:ins}
\end{table}

For all the instances, each integer and real variable was standardized so that its mean was zero and its standard deviation was one. 
Each categorical variable was transformed into dummy variable(s). 
Variables having missing values for samples of over 10\% were eliminated. 
After that, samples including missing values were all eliminated.
In the \texttt{Tnns-W} and \texttt{Tnns-M} instances, the variables ``Player 1'' and ``Player 2'' were removed because they are not suitable for prediction purposes. 

The computational experiments compared the performances of the following methods: 
\begin{description}
\item[Quad] MIQO formulation~\eqref{MIQO:obj}--\eqref{MIQO:const3} based on quadratic approximation, 
\item[PWL] MILO formulation~\eqref{MILO:obj}--\eqref{MILO:const3} based on piecewise-linear approximation using the following set of points for tangent lines, similarly to Sato et al.~\cite{SaTa15}:
\[
V = \{0, \pm 0.44, \pm 0.89, \pm 1.37, \pm 1.90, \pm 2.63, \pm 3.55, \pm 5.16, \pm \infty\} \quad (|V| = 17). 
\]
\end{description}

All computations were performed on a Linux computer with an Intel Core i7-4820 CPU
(3.70 GHz) and 32 GB memory. 
Gurobi Optimizer 6.0.0~(\url{http://www.gurobi.com}) was used to solve the MILO and MIQO problems. 
Here, the logical implications~\eqref{MIQO:const1}~and~\eqref{MILO:const2} were represented by the {\tt SOS\_TYPE1} function in Gurobi Optimizer. 

Tables~\ref{tab:aic_f}--\ref{tab:bic_b} show the computational results of minimizing AIC/BIC in the forward/backward sequential logit models. 
The columns labeled ``AIC'' and ``BIC'' are the values of the corresponding information criteria calculated from the selected set of features. 
Note that the smaller of the AIC/BIC values between Quad and PWL are bold-faced for each instance. 
The column labeled ``objval'' is the value of the objective function, i.e., \eqref{MIQO:obj} and \eqref{MILO:obj}. 
The column labeled ``$|S|$'' is the number of selected features, and the column labeled ``time~(s)'' is computation time in seconds. 
The computation for solving the MILO/MIQO problems was terminated if it did not finish by itself after 10000 seconds. 
In this case, the tables show the result of the best solution obtained within 10000 seconds. 

\begin{table}[!t]
\begin{tabular}{c}
\begin{minipage}{1\hsize}
\centering
\caption{AIC minimization in forward sequential logit model}
\begin{tabular}{lrrrlrrrr}
\toprule
instance & $n$ & $p$ & \#class & method & AIC & objval & $|S|$ & time~(s) \\ \hline
\texttt{Wine-R}& 1599 & 11 & 6 &
     Quad & 3057.5 & 4204.6 & 4 & 0.03 \\
&&&& PWL & \textbf{3028.4} & 3013.2 & 10 & 428.05 \\ \hline
\texttt{Wine-W}& 4898 & 11 & 7 &
     Quad & 10859.6 & 14343.0 & 8 & 0.07  \\
&&&& PWL & \textbf{10726.7} & 10671.2 & 9 & 1899.54 \\ \hline
\texttt{Skill} & 3338 & 18 & 7 &
     Quad & 9108.8 & 11289.1 & 9 & 5.13 \\
&&&& PWL & \textbf{9080.2} & 8939.0 & 15 & $>$10000 \\ \hline
\texttt{Choice}   & 1474 & 21 & 3 &
     Quad & 2816.2 & 2850.2 & 9 & 7.07 \\
&&&& PWL & \textbf{2813.1} & 2804.4 & 12 & 1632.37 \\ \hline
\texttt{Tnns-W}& 118 & 31 & 7 &
     Quad & 331.1 & 331.1 & 0 & $>$10000 \\
&&&& PWL & \textbf{316.2} & 315.4 & 4 & $>$10000 \\ \hline
\texttt{Tnns-M}& 113 & 33 & 7 &
     Quad & 278.5 & 296.1 & 2 & $>$10000 \\
&&&& PWL & \textbf{278.3} & 277.6 & 4 & $>$10000 \\ \hline
\texttt{Stdnt-M} & 395  & 40 & 18 &
     Quad & 1052.7 & 2251.2 & 1 & $>$10000 \\
&&&& PWL & \textbf{946.2} & 941.9 & 6 & $>$10000 \\ \hline
\texttt{Stdnt-P} & 649  & 40 & 17 &
     Quad & 1709.9 & 3929.7 & 1 & $>$10000 \\
&&&& PWL & \textbf{1653.6} & 1645.2 & 7 & $>$10000 \\
\bottomrule
\end{tabular}
\label{tab:aic_f}
\end{minipage}
\\\\
\begin{minipage}{1\hsize}
\centering
\caption{AIC minimization in backward sequential logit model}
\begin{tabular}{lrrrlrrrr}
\toprule
instance & $n$ & $p$ & \#class & method & AIC & objval & $|S|$ & time~(s) \\ \hline
\texttt{Wine-R}& 1599 & 11 & 6 &
     Quad & 3062.5 & 4073.5 & 5 & 0.04 \\
&&&& PWL & \textbf{3050.1} & 3034.6 & 10 & 357.76 \\ \hline
\texttt{Wine-W}& 4898 & 11 & 7 &
     Quad & 10811.6 & 14697.4 & 7 & 0.05 \\
&&&& PWL & \textbf{10786.9} & 10734.7 & 9 & 1785.95 \\ \hline
\texttt{Skill} & 3338 & 18 & 7 &
     Quad & 9024.5 & 11089.6 & 8 &  22.65 \\
&&&& PWL & \textbf{8961.2} & 8925.9 & 10 & $>$10000 \\ \hline
\texttt{Choice}   & 1474 & 21 & 3 &
     Quad & 2829.4 & 2940.8 & 9 & 11.51 \\
&&&& PWL & \textbf{2826.1} & 2816.7 & 11 & 3513.40 \\ \hline
\texttt{Tnns-W}& 118 & 31 & 7 &
     Quad & 331.1 & 447.4 & 0 & 253.13 \\
&&&& PWL & \textbf{327.0} & 307.5 & 8 & $>$10000 \\ \hline
\texttt{Tnns-M}& 113 & 33 & 7 &
     Quad & 307.4 & 419.8 & 0 & 9.53 \\
&&&& PWL & \textbf{280.5} & 279.3 & 4 & $>$10000 \\ \hline
\texttt{Stdnt-M} & 395  & 40 & 18 &
     Quad & 1065.5 & 2584.6 & 2 & $>$10000 \\
&&&& PWL & \textbf{1044.1} & 1038.1 & 2 & $>$10000 \\ \hline
\texttt{Stdnt-P} & 649  & 40 & 17 &
     Quad & 1640.7 & 3532.1 & 2 & $>$10000 \\
&&&& PWL & \textbf{1620.7} & 1611.6 & 3 & $>$10000  \\
\bottomrule
\end{tabular}
\label{tab:aic_b}
\end{minipage}
\end{tabular}
\end{table}

\begin{table}[!t]
\begin{tabular}{c}
\begin{minipage}{1\hsize}
\centering
\caption{BIC minimization in forward sequential logit model}
\begin{tabular}{lrrrlrrrr}
\toprule
instance & $n$ & $p$ & \#class & method & BIC & objval & $|S|$ & time~(s) \\ \hline
\texttt{Wine-R}& 1599 & 11 & 6 &
     Quad & \textbf{3191.9} & 4339.1 & 4 & 0.05 \\
&&&& PWL & \textbf{3191.9} & 3175.3 & 4 & 331.58 \\ \hline
\texttt{Wine-W}& 4898 & 11 & 7 &
     Quad & 11127.4 & 14543.6 & 3 & 0.05 \\
&&&& PWL & \textbf{11077.0} & 11019.4 & 4 & 6901.04 \\ \hline
\texttt{Skill} & 3338 & 18 & 7 &
     Quad & 9434.7 & 11518.1 & 3 & 1.41 \\
&&&& PWL & \textbf{9333.3} & 9291.3 & 6 & $>$10000 \\ \hline
\texttt{Choice}   & 1474 & 21 & 3 &
     Quad & \textbf{2903.4} & 2932.4 & 5 & 3.00 \\
&&&& PWL & \textbf{2903.4} & 2894.5 & 5 & 1372.28 \\ \hline
\texttt{Tnns-W}& 118 & 31 & 7 &
     Quad & \textbf{347.7} & 347.7 & 0 & 99.67 \\
&&&& PWL & \textbf{347.7} & 345.9 & 0 & 833.61 \\ \hline
\texttt{Tnns-M}& 113 & 33 & 7 &
     Quad & \textbf{323.7} & 323.7 & 0 & 96.83 \\
&&&& PWL & \textbf{323.7} & 321.3 & 0 & 514.01 \\ \hline
\texttt{Stdnt-M} & 395  & 40 & 18 &
     Quad & 1187.9 & 2386.4 & 1 & 220.77 \\
&&&& PWL & \textbf{1181.7} & 1175.9 & 2 & $>$10000 \\ \hline
\texttt{Stdnt-P} & 649  & 40 & 17 &
     Quad & \textbf{1853.2} & 4073.0 & 1 & 725.73 \\
&&&& PWL & \textbf{1853.2} & 1835.3 & 1 & $>$10000 \\
\bottomrule
\end{tabular}
\label{tab:bic_f}
\end{minipage}
\\\\
\begin{minipage}{1\hsize}
\centering
\caption{BIC minimization in backward sequential logit model}
\begin{tabular}{lrrrlrrrr}
\toprule
instance & $n$ & $p$ & \#class & method & BIC & objval & $|S|$ & time~(s) \\ \hline
\texttt{Wine-R}& 1599 & 11 & 6 &
     Quad & 3257.3 & 4197.6 & 2 & 0.02 \\
&&&& PWL & \textbf{3206.5} & 3190.6 & 4 & 296.37 \\ \hline
\texttt{Wine-W}& 4898 & 11 & 7 &
     Quad & 11151.9 & 14907.6 & 3 & 0.03 \\
&&&& PWL & \textbf{11143.2} & 11083.8 & 4 & 8233.98 \\ \hline
\texttt{Skill} & 3338 & 18 & 7 &
     Quad & 9425.3 & 11315.8 & 4 & 3.47 \\
&&&& PWL & \textbf{9275.4} & 9241.3 & 6 & $>$10000 \\ \hline
\texttt{Choice}   & 1474 & 21 & 3 &
     Quad & 2917.3 & 3009.5 & 4 & 1.86 \\
&&&& PWL & \textbf{2917.0} & 2907.5 & 5 & 1601.25 \\ \hline
\texttt{Tnns-W}& 118 & 31 & 7 &
     Quad & \textbf{347.7} & 464.0 & 0 & 0.10 \\
&&&& PWL & \textbf{347.7} & 344.1 & 0 & 3205.70 \\ \hline
\texttt{Tnns-M}& 113 & 33 & 7 &
     Quad & \textbf{323.7} & 436.1 & 0 & 0.09 \\
&&&& PWL & \textbf{323.7} & 319.9 & 0 & 9767.70 \\ \hline
\texttt{Stdnt-M} & 395  & 40 & 18 &
     Quad & \textbf{1207.0} & 2740.4 & 1 & 421.56 \\
&&&& PWL & \textbf{1207.0} & 1199.8 & 1 & $>$10000 \\ \hline
\texttt{Stdnt-P} & 649  & 40 & 17 &
     Quad & \textbf{1822.4} & 3717.1 & 1 & $>$10000 \\
&&&& PWL & \textbf{1822.4} & 1809.3 & 1 & $>$10000 \\
\bottomrule
\end{tabular}
\label{tab:bic_b}
\end{minipage}
\end{tabular}
\end{table}

Tables~\ref{tab:aic_f}~and~\ref{tab:aic_b} show the results of AIC minimization in the forward and backward sequential logit models. 
These tables reveal that AIC values of PWL were smaller than those of Quad for all the instances. 
The computation time of PWL was longer than that of Quad because the problem size of PWL is dependent on the number of samples (see the constraints~\eqref{MILO:const1}). 
Nevertheless, we should notice that PWL provided better-quality solutions than Quad did in spite of the time limit of 10000 seconds. 
In addition, the number of features selected by PWL sometimes differed greatly from that selected by Quad; for instance, Quad and PWL respectively selected 4 and 10 features for \texttt{Wine-R} in Table~\ref{tab:aic_f}. 

We can see from Tables~\ref{tab:aic_f}~and~\ref{tab:aic_b} that PWL approximated the logistic loss function very accurately, whereas Quad caused a major gap between AIC and objval. 
Additionally, since the objective function of PWL is an underestimator relative to AIC, its optimal value serves as a lower bound of the smallest AIC. 
In the case of \texttt{Wine-R} in Table~\ref{tab:aic_f}, AIC and objval of Quad were 3057.5 and 4204.6, whereas those of PWL were 3028.4 and 3013.2. 
This also implies that PWL found a subset of features such that the associated AIC value was 3028.4, and it collaterally guaranteed that the smallest AIC value was greater than 3013.2. 
This optimality guarantee is the most notable characteristic of our piecewise-linear approximation approach, and it cannot be shared by the quadratic approximation approach. 

Tables~\ref{tab:bic_f}~and~\ref{tab:bic_b} show the results of BIC minimization in the forward and backward sequential logit models. 
We should recall that the penalty, $F$, for the number of selected features in BIC (i.e., $F = \log(n)$) is larger than that in AIC (i.e., $F = 2$). 
Hence, BIC minimization selected a small number of features, and accordingly, Quad and PWL often yielded the same subset of features for each instance in Tables~\ref{tab:bic_f}~and~\ref{tab:bic_b}. 
Meanwhile, when these two methods provided different subsets of features, PWL always found the better one. 

\section{Conclusions}
This paper dealt with the feature subset selection problem for a sequential logit model. 
We formulated it as a mixed integer linear optimization (MILO) problem by applying a piecewise-linear approximation to the logistic loss functions. 
The computational results confirmed that our formulation has a clear advantage over the mixed integer quadratic optimization (MIQO) formulation proposed in the previous study~\cite{TaNa14}. 

In contrast to the MIQO formulation, the approximation accuracy of the logistic loss function can be controlled by the number of tangent lines in our MILO formulation. 
Furthermore, after the MILO problem is solved, it provides an optimality guarantee of the selected features on the basis of information criteria. 
To the best of our knowledge, this paper is the first to compute a subset of features with an optimality guarantee for a sequential logit model. 

A future direction of study will be to extend our piecewise-linear approximation approach to other logit models. 
However, this will be a difficult task because it is imperative to approximate a multivariate objective function. 
Another direction of future research is to analyze actual data by means of our feature subset selection method. 
For instance, Sato et al.~\cite{SaTa} investigated consumers' store choice behavior by applying feature subset selection based on mixed integer optimization. 
Since proper feature subset selection is essential for data analysis, our approach has a clear advantage over heuristic algorithms. 


\section*{Acknowledgments}
This work was partially supported by Grants-in-Aid for Scientific
Research by the Ministry of Education, Culture, Sports, Science and
Technology of Japan.


\vspace{15pt}
{\leftskip 6cm \parindent 0mm
Toshiki Sato\\
Doctoral Program in Social Systems and Management \\
Graduate School of Systems and Information Engineering \\
University of Tsukuba \\
1-1-1 Tennodai, Tsukuba-shi, Ibaraki 305-8573, Japan \\ 
E-mail: \texttt{tsato@sk.tsukuba.ac.jp}
\par}

\begin{thebibliography}{99}
\bibitem{AfGo09}
A. Afonso, P. Gomes, and P. Rother: 
Ordered response models for sovereign debt ratings. 
\textit{Applied Economics Letters,} \textbf{16} (2009), 769--773.

\bibitem{Ag90}
A. Agresti:
\textit{Categorical Data Analysis} (Wiley, New York, 1990).

\bibitem{Ag10}
A. Agresti:
\textit{Analysis of Ordinal Categorical Data, Second Edition} (John Wiley \& Sons, New York, 2010).

\bibitem{Ak74}
H. Akaike:
A new look at the statistical model identification.
\textit{IEEE Transactions on Automatic Control,} \textbf{19} (1974), 716--723.

\bibitem{Am81}
T. Amemiya:
Qualitative response models: A survey. 
\textit{Journal of Economic Literature,} \textbf{19} (1981), 1483--1536.

\bibitem{Ar12}
K. J. Archer and A. A. A. Williams: 
$L_1$ penalized continuation ratio models for ordinal response prediction using high-dimensional datasets. 
\textit{Statistics in Medicine,} \textbf{31} (2012), 1464--1474.

\bibitem{BaLi13}
K. Bache and M. Lichman:
UCI Machine Learning Repository [\url{http://archive.ics.uci.edu/ml}].
Irvine, CA: University of California, School of Information and Computer Science (2013).

\bibitem{Be63}
E. M. L. Beale:
Two transportation problems. 
In G. Kreweras and G. Morlat (eds.):
\textit{Proceedings of the Third International Conference on Operational Research} (Dunod, Paris and English Universities Press, London, 1963), 780--788.

\bibitem{BeTo70}
E. M. L. Beale and J. A. Tomlin:
Special facilities in a general mathematical programming system for non-convex problems using ordered sets of variables. 
In J. Lawrence (ed.):
\textit{Proceedings of the Fifth International Conference on Operational Research} (Tavistock Publications, London, 1960), 447--454.

\bibitem{BeKi15}
D. Bertsimas, A. King, and R. Mazumder:
Best subset selection via a modern optimization lens. 
arXiv preprint arXiv:1507.03133 (2015).

\bibitem{BlLa97}
A. L. Blum and P. Langley:
Selection of relevant features and examples in machine learning. 
\textit{Artificial Intelligence,} \textbf{97} (1997), 245--271.

\bibitem{Ed10}
L. H. Ederington:
Classification models and bond ratings. 
\textit{Financial Review,} \textbf{20} (1985), 237--262.

\bibitem{Ef60}
M. A. Efroymson:
Multiple regression analysis. 
In A. Ralston and H. S. Wilf (eds.):
\textit{Mathematical Methods for Digital Computers} (Wiley, New York, 1960) 191--203.

\bibitem{Fi80}
S. E. Fienberg:
\textit{The Analysis of Cross-Classified Categorical Data}
(The MIT Press, Cambridge, 1980)

\bibitem{FuWi04}
H. Fu and C. Wilmot:
Sequential logit dynamic travel demand model for hurricane evacuation. 
\textit{Transportation Research Record: Journal of the Transportation Research Board,} \textbf{1882} (2004), 19--26.

\bibitem{GuEl03}
I. Guyon and A. Elisseeff:
An introduction to variable and feature selection. 
\textit{The Journal of Machine Learning Research,} \textbf{3} (2003), 1157--1182.

\bibitem{KaMo79}
L. M. Kahn and K. Morimune:
Unions and employment stability: A sequential logit approach. 
\textit{International Economic Review,} \textbf{20} (1979), 217--235.

\bibitem{Ka09}
L. N. Kazembe:
A semiparametric sequential ordinal model with applications to analyse first birth intervals. 
\textit{Austrian Journal of Statistics,} \textbf{38} (2009), 83--99.

\bibitem{Ki08}
H. T. Kiiveri:
A general approach to simultaneous model fitting and variable elimination in response models for biological data with many more variables than observations. 
\textit{BMC Bioinformatics,} \textbf{9} (2008), 195.

\bibitem{KoJo97}
R. Kohavi and G. H. John:
Wrappers for feature subset selection. 
\textit{Artificial Intelligence,} \textbf{97} (1997), 273--324.

\bibitem{KoTa10}
H. Konno and Y. Takaya:
Multi-step methods for choosing the best set of variables in regression analysis. 
\textit{Computational Optimization and Applications,} \textbf{46} (2010), 417--426.

\bibitem{KoYa09}
H. Konno and R. Yamamoto:
Choosing the best set of variables in regression analysis using integer programming. 
\textit{Journal of Global Optimization,} \textbf{44} (2009), 273--282.

\bibitem{LiAg05}
I. Liu and A. Agresti:
The analysis of ordered categorical data: An overview and a survey of recent developments. 
\textit{Test,} \textbf{14} (2005), 1--73.

\bibitem{Mi02}
A. Miller:
\textit{Subset Selection in Regression, Second Edition}
(CRC Press, Boca Raton, 2002).

\bibitem{MiTa15a}
R. Miyashiro and Y. Takano:
Subset selection by Mallows' $C_p$: A mixed integer programming approach. 
\textit{Expert Systems with Applications,} \textbf{42} (2015), 325--331.

\bibitem{MiTa15b}
R. Miyashiro and Y. Takano:
Mixed integer second-order cone programming formulations for variable selection in linear regression. 
\textit{European Journal of Operational Research,} \textbf{247} (2015), 721--731. 

\bibitem{NaKo09}
D. Nagakura and M. Kobayashi:
Testing the sequential logit model against the nested logit model. 
\textit{Japanese Economic Review,} \textbf{60} (2009), 345--361.

\bibitem{PoFi99}
W. P. Poon, M. Firth, and H. G. Fung:
A multivariate analysis of the determinants of Moody's bank financial strength ratings. 
\textit{Journal of International Financial Markets, Institutions and Money,} \textbf{9} (1999), 267--283.


\bibitem{SaTa15}
T. Sato, Y. Takano, R. Miyashiro, and A. Yoshise:
Feature subset selection for logistic regression via mixed integer optimization.
Department of Policy and Planning Sciences, Discussion Paper Series, No. 1324. University of Tsukuba (2015).

\bibitem{SaTa}
T. Sato, Y. Takano, and T. Nakahara: 
Using mixed integer optimization to select variables for a store choice model. 
\textit{International Journal of Knowledge Engineering and Soft Data Paradigms} (in press). 

\bibitem{Sc78}
G. Schwarz:
Estimating the dimension of a model. 
\textit{The Annals of Statistics,} \textbf{6} (1978), 461--464.

\bibitem{TaNa14}
K. Tanaka and H. Nakagawa:
A method of corporate credit rating classification based on support vector machine and its validation in comparison of sequantial logit model.
\textit{Transactions of the Operations Research of Japan,} \textbf{57} (2014), 92--111.

\bibitem{Tu91}
G. Tutz:
Sequential models in categorical regression. 
\textit{Computational Statistics \& Data Analysis,} \textbf{11} (1991), 275--295.

\bibitem{Tu10}
G. Tutz and A. Groll:
Binary and ordinal random effects models including variable selection. 
Technical Report Number 097, 2010. Department of Statistics University of Munich (2010).

\bibitem{We86}
W. C. Weiler:
A sequential logit model of the access effects of higher education institutions. 
\textit{Economics of Education Review,} \textbf{5} (1986), 49--55.

\bibitem{Yu09}
S. C. Yusta:
Different metaheuristic strategies to solve the feature selection problem. 
\textit{Pattern Recognition Letters,} \textbf{30} (2009), 525--534.

\end{thebibliography}
\end{document}